\documentstyle[12pt]{article}
\title{\bf Euclidean wormholes with Phantom field and Phantom field
accompanied by perfect fluid}
\author{F. Darabi \thanks{f.darabi@azaruniv.edu} \\
{\small Department of Physics, Azarbaijan University of Tarbiat
Moallem, 53714-161 Tabriz, Iran}}
\begin{document}
\maketitle
\begin{abstract}
We study the classical Euclidean wormhole solutions for the gravitational
systems with minimally coupled pure Phantom field and minimally coupled  Phantom field accompanied by perfect fluid. It is shown that such
solutions do exist and then the general forms of the Phantom field potential are obtained for which there are classical Euclidean wormhole solutions. 
\end{abstract}

\newpage

\section{Introduction}

Classical wormholes are usually considered as Euclidean metrics
that consist of two asymptotically flat regions connected by a
narrow throat (handle). Wormholes have been studied mainly as
instantons, namely solutions of the classical Euclidean field
equations. In general, such wormholes can represent quantum
tunneling between different topologies. They are possibly useful
in understanding black hole evaporation \cite{Haw}; in allowing
nonlocal connections that could determine fundamental constants;
and in vanishing the cosmological constant $\Lambda$ \cite{Col1}-\cite{Col3}.
They are even considered as an alternative to the Higgs mechanism
\cite{Mig}. Consequently, such solutions are worth finding.

The reason why classical wormholes may exist is related to the
implication of a theorem of Cheeger and Glommol \cite{Chee} which
states that a {\it necessary} (but not sufficient) condition for a classical wormhole to exist is that the eigenvalues of the Ricci tensor be negative {\it somewhere} on the manifold. Unfortunately, there exists certain special kinds of matter for which the above necessary condition is satisfied. 
For example, the energy-momentum tensors of an axion field and of a conformal
scalar field are such that, when coupled to gravity, the Ricci
tensor has negative eigenvalues. However, for pure gravity or a (real) minimally coupled scalar field it is easy to show that the Ricci tensor can never be negative. It is shown that the (pure imaginary) minimally coupled scalar field, conformally coupled scalar field, a third rank antisymmetric tensor field and some special kinds of matter sources have wormhole solutions \cite{Gid1}-\cite{Gid5}.

In the present paper we shall investigate the possibility for the existence of classical Euclidean wormholes in the gravitational models minimally coupled
with the pure Phantom field, and Phantom field accompanied by perfect fluid. Historically, Phantom fields were first introduced in Hoyle's version of the steady state theory. In adherence
to the perfect cosmological principle, a creation field (Cfield)
was introduced by Hoyle to reconcile the model with the homogeneous density of the universe by the creation of new matter in the voids caused by the expansion of the universe \cite{Hoyl}. It was further refined and reformulated
in the Hoyle and Narlikar theory of gravitation \cite{HN}. From the cosmological point of view, the Phantom field is a good candidate for exotic matter which is characterized by the equation of state $\omega = {p}/{\rho} < -1$. Specific models in braneworlds or Brans-Dicke scalar-tensor gravity can lead to Phantom energy \cite{Sahni}. Meanwhile the simplest
explanation for the Phantom dark energy is provided
by a scalar field with a negative kinetic energy \cite{Cald} .
Such a field may be motivated from S-brane constructions
in string theory \cite{Sbrane}. In section 2, a brief review is given for the non-existence of Euclidean wormholes in the gravitational model minimally coupled with the real scalar field. In section 3, we introduce the action for the Phantom field minimally coupled with gravity and obtain the corresponding Ricci tensor to show the existence of classical Euclidean wormholes. Then, the explicit form of the potential is obtained for which Euclidean wormholes do exist. In section 4, we study the action for the Phantom field accompanied by perfect fluid minimally coupled with gravity and obtain the corresponding Ricci tensor and then show the existence of classical Euclidean wormholes in this model as well. The explicit forms of the potential is then obtained for which Euclidean wormholes exist. The paper is ended with a conclusion.

\section{No wormholes with minimally coupled real scalar field}

It is well known that classical Euclidean wormholes can occur if the Ricci
tensor has negative eigenvalues somewhere on the manifold. Actually, this
is necessary but not sufficient condition for their existence
and is related to the implication of a theorem of Cheeger and Glommol \cite{Chee}.
For example, the energy-momentum tensors of an axion field and of a conformal
scalar field are such that, when coupled to gravity, the Ricci
tensor has negative eigenvalues. However, the story of minimally coupled real scalar field was somehow controversial. Costakis {\it et al} \cite{Cost}, had already claimed that Euclidean wormholes could be obtained with minimally
coupled real scalar field, in contradiction with widely belief that wormholes
require rather unusual matter sources \cite{HP}. However, Coule commented
on this claim and showed that an imaginary scalar field is necessary for Euclidean
wormholes to occur \cite{Coule}. 

Let us consider the action of an ordinary scalar field minimally coupled to gravity\footnote{We use the units in which $\frac{8\pi G}{3}=1$.}\cite{Sahni1}
\begin{equation}\label{1}
{\cal S}=\int d^4x \sqrt{-g}\left[{\cal R}-\frac{1}{2}\nabla_{\mu}\phi \nabla^{\mu}\phi-V(\phi)\right],
\end{equation}
where $V(\phi)$ is the potential of the scalar field. 
The energy momentum tensor of the field is derived by
varying the action in terms of $g^{\mu \nu}$
\begin{eqnarray}\label{2}
T_{\mu \nu}&=&-\frac{2}{\sqrt{-g}}\frac{\delta{\cal S}}{\delta g^{\mu \nu}}\\ \nonumber
&=&\nabla_{\mu}\phi \nabla_{\nu}\phi-g_{\mu
\nu}\left[\frac{1}{2}\nabla_{\lambda}\phi\nabla^{\lambda}\phi+V(\phi)\right].
\end{eqnarray}
The trace is also obtained as 
\begin{equation}\label{3}
T=-\nabla_{\mu}\phi \nabla^{\mu}\phi-4V(\phi).
\end{equation}
The Einstein equations 
\begin{equation}\label{4}
R_{\mu \nu}=T_{\mu \nu}-\frac{1}{2}g_{\mu \nu}T,
\end{equation}
give
\begin{equation}\label{5}
R_{\mu \nu}=\nabla_{\mu}\phi \nabla_{\nu}\phi+ g_{\mu \nu}V(\phi),
\end{equation}
which shows $R_{\mu \nu}$ can never be negative in Euclidean space $g_{\mu \nu}=(++++)$ for $V(\phi)\geq 0$ unless we consider
a purely imaginary scalar field, i.e., let $\phi \rightarrow i\phi$ \cite{Coule}.
We consider the Friedmann-Robertson-Walker (FRW) metric in its Euclidean form 
\begin{equation}
ds^2=d\tau^2+a^2(\tau)\left(\frac{dr^2}{(1-kr^2)}+r^2(d\theta^2+\sin^2\theta
d\phi^2 )\right),
\end{equation}
where $k=+1,0, -1$ denotes the closed, flat and open universes, respectively. The field equations for the closed universe are obtained 
\begin{equation}\label{6}
\frac{\dot{a}^2}{a^2}=\frac{1}{a^2}+\left[\frac{1}{2}\dot{\phi}^2-V(\phi)\right],
\end{equation}
\begin{equation}\label{7}
\ddot{\phi}+3\frac{\dot{a}}{a}\dot{\phi}=\frac{dV(\phi)}{d\phi},
\end{equation}
where a dot denotes derivatives with respect to Euclidean time $\tau$. By
taking the ansatz \cite{Cost}
\begin{equation}\label{8}
\dot{\phi}=\frac{A}{a^p},
\end{equation}
where $A$ is a constant and $p$ is a positive integer, the potential takes
on the following form
\begin{equation}\label{9}
V(\phi(a))=\frac{A^2(p-3)}{2pa^{2p}}.
\end{equation}
Substituting this form into Eq.(\ref{6}) gives
\begin{equation}\label{10}
\frac{\dot{a}^2}{a^2}=\frac{1}{a^2}+\frac{3A^2}{2pa^{2p}}.
\end{equation}
In spatially closed FRW models, wormholes are typically described by a constraint equation of the form
\begin{equation}\label{11}
\frac{\dot{a}^2}{a^2}=\frac{1}{a^2}-\frac{Const}{a^n}.
\end{equation}
In order an asymptotically Euclidean wormhole to exist it is necessary that
$\dot{a}^2>0$ for large $a$ which requires $n>2$. Then, this wormhole has two asymptotically flat
regions connected by a throat where $\dot{a}=0$. Comparing Eq.(\ref{10})
with Eq.(\ref{11}) reveals that the real scalar field can not represent Euclidean
wormholes and in order to have Euclidean wormholes we should let $A \rightarrow i A$ (or $\phi \rightarrow i \phi$) so that
\begin{equation}\label{12}
\frac{\dot{a}^2}{a^2}=\frac{1}{a^2}-\frac{3A^2}{2pa^{2p}},
\end{equation}
which now agrees with the typical known wormholes such as $p=2$ or $n=4$
(conformal scalar field \cite{Gid2}) and $p=3$ or $n=6$
(axion field \cite{Gid1}).

\section{Euclidean wormholes with minimally coupled pure Phantom field}

In the previous section we realized that the negative eigenvalues of Ricci
tensor for an imaginary scalar field is caused, in principle, by the the derivative part alone being negative definite, although the potential is
also affected by $\phi \rightarrow i \phi$. However, one may consider a model
of real scalar field whose kinetic term is intrinsically negative definite. This is the so called {\it Phantom} field, and we will show that the wormhole solutions like those obtained by Costakis {\it et al} hold for this model.
\\
The action for the Phantom field minimally coupled to gravity is written \cite{Sahni1}
\begin{equation}\label{13}
{\cal S}=\int d^4x \sqrt{-g}\left[{\cal R}+\frac{1}{2}\nabla_{\mu}\phi \nabla^{\mu}\phi-V(\phi)\right].
\end{equation}
The energy momentum tensor and its trace are given by
\begin{eqnarray}\label{14}
T_{\mu \nu}=-\nabla_{\mu}\phi \nabla_{\nu}\phi+g_{\mu
\nu}\left[\frac{1}{2}\nabla_{\lambda}\phi\nabla^{\lambda}\phi-V(\phi)\right],
\end{eqnarray}
\begin{equation}\label{15}
T=\nabla_{\mu}\phi \nabla^{\mu}\phi-4V(\phi).
\end{equation}
The Einstein equations read
\begin{equation}\label{16}
R_{\mu \nu}=-\nabla_{\mu}\phi \nabla_{\nu}\phi+ g_{\mu \nu}V(\phi).
\end{equation}
It turns out that $R_{\mu \nu}$ may be negative in Euclidean space $g_{\mu \nu}=(++++)$ if and only if the following condition is satisfied
\begin{equation}\label{17}
-\nabla_{\mu}\phi \nabla_{\nu}\phi+ g_{\mu \nu}V(\phi)<0,
\end{equation}
where the potential may take one of the following cases
\begin{equation}\label{18}
 \left\{ \begin{array}{ll} V(\phi)\geq 0, 
 
 \\V(\phi)< 0.
 \\
\end{array}
\right.
\end{equation}
The field equations for the closed universe are obtained 
\begin{equation}\label{19}
\frac{\dot{a}^2}{a^2}=\frac{1}{a^2}-\left[\frac{1}{2}\dot{\phi}^2+V(\phi)\right],
\end{equation}
\begin{equation}\label{20}
\ddot{\phi}+3\frac{\dot{a}}{a}\dot{\phi}=-\frac{dV(\phi)}{d\phi}.
\end{equation}
By choosing the ansatz (\ref{8}) the potential casts
in the following form
\begin{equation}\label{21}
V(\phi(a))=\frac{A^2(3-p)}{2pa^{2p}},
\end{equation}
which after substituting into Eq.(\ref{19}) gives
\begin{equation}\label{22}
\frac{\dot{a}^2}{a^2}=\frac{1}{a^2}-\frac{3A^2}{2pa^{2p}},
\end{equation}
which is in agreement with the typical wormholes equation (\ref{11}) provided
that $p>1$. 

It is now important to explain about and compare the results we have so far obtained and those obtained in \cite{Cost}. At first sight, comparing Eqs.(12) and (13) in \cite{Cost} with (\ref{19}) and (\ref{20}) here, it seems that the only difference is a missing minus sign of the kinetic term in the comparison of Friedmann equations (12) and (\ref{19}) which seems to be corrected by assuming a map $\phi \rightarrow i\phi$. However, the point is that this map affects the potential as well and the equations become different again. In fact,
as explained in \cite{Coule}, the major cause for the missing minus sign of the kinetic term in \cite{Cost} arises from the use of a different convention for the energy-momentum energy tensor and a crucial mistake about a + sign in front of the potential in Eq.(12). In the present paper, similar to \cite{Coule}
we have used the conventions of Wald \cite{Wald} for the energy-momentum
tensor of the scalar field which provides us with the correct forms of Eqs.(\ref{19}) and (\ref{20}) which result in Eqs.(\ref{21}) and (\ref{22}). 

Now, we may realize the important differences between the results obtained
here and those of obtained in \cite{Cost}. The equation (\ref{21}) is the same as the one obtained in \cite{Cost}, but the equation (\ref{22}) is different from the corresponding one in \cite{Cost} in that the functional dependence of ${\dot{a}^2}/a^2$ on $p$ has now been changed. Moreover, the one condition (9) in \cite{Cost} is replaced here by the two conditions considered in (\ref{18}). This is because the missing minus sign in Eq.(8) of Ref.\cite{Cost} requires
mistakenly that the potential must always be positive, whereas in the correct
form of this equation for the case of Phantom field, namely (\ref{16}), the
potential can take the negative values as well. Therefore, the results obtained in \cite{Cost} are just the $V(\phi)\geq0$ sector of the general conditions (\ref{18}) which we shall study in the following cases.

\subsection{Case: $V(\phi)\geq0$ }

It is clear that in order the potential (\ref{21}) satisfies the condition
$V(\phi)\geq0$ it is necessary that $p\leq3$. We then consider the three cases for $p$ as follows.

\subsubsection{\bf Case  $p=3$: }

This case corresponds to a vanishing potential but 
the scalar field is still dynamical, namely $\dot{\phi}\neq 0$. The Friedmann
equation (\ref{22}) is written
\begin{equation}\label{29}
\frac{\dot{a}^2}{a^2}=\frac{1}{a^2}-\frac{A^2}{2a^{6}},
\end{equation}
which is in the form of Euclidean wormhole (\ref{11}). This equation solves
to \cite{Cost}
\begin{equation}\label{30}
\frac{1}{\sqrt{2}}F(\delta,r)-\sqrt{2}E(\delta,r)+\frac{1}{u}\sqrt{u^2-1}=\frac{2^{1/4}}{\sqrt{-A}}(\tau-\tau_0),
\end{equation}
where $F(\delta,r)$ and $E(\delta,r)$ are the elliptic integrals of the first
and second kind, respectively, with $\delta=\arccos(1/u)$, $r=1/\sqrt{2}$,
and $u=(2^{1/4}/\sqrt{-A})a$. By inverting Eq.(\ref{30}) the corresponding
form of $a(\tau)$ is obtained.

\subsubsection{\bf Case  $p=2$:}

In this case, using the same ansatz (\ref{8}) the system of equations (\ref{19}), (\ref{20}) leads to
\begin{equation}\label{26}
a^2(\tau)=\frac{3A^2}{4}+(\tau-c_1)^2,
\end{equation}
\begin{equation}\label{27}
\phi(\tau)=\frac{2}{\sqrt{3}}\arctan\left({\frac{2}{A\sqrt{3}}(\tau-c_1)}\right)+c_2,
\end{equation}
\begin{equation}\label{28}
V(\phi)= \frac{4}{9A^2}\cos^4\left(\frac{\sqrt{3}(\phi-c_2)}{2}\right).
\end{equation}
It turns out that (\ref{26}) represents a wormhole solution with the minimum
radius (throat) of the size $a_{min}\sim |A|$ which is connected to asymptotically
flat space as $\tau \rightarrow \pm \infty$.

\subsubsection{\bf Case  $p=1$:}

The system of equations (\ref{19}), (\ref{20}) for the
ansatz (\ref{8}) leads to \cite{Cost}
\begin{equation}\label{23}
a(\tau)=a_0+a_1\tau,
\end{equation}
\begin{equation}\label{24}
\phi(\tau)=\frac{A}{a_1}\ln(a_0+a_1\tau)+a_2,
\end{equation}
\begin{equation}\label{25}
V(\phi)\sim \exp(\phi-a_3),
\end{equation}
where $a_0, a_1, a_2,$ and $a_3$ are constants of integrations. The solution
(\ref{23}) does not show a wormhole. This is because the {\it boundary} condition $\dot{a}=0$ kills the coefficient $a_1$ and we loose the time dependence
of the solution. Moreover, the radius $a$ becomes negative for $\tau \rightarrow
-\infty$. Actually, the reason why this case does not lead to a wormhole
is simple: the cases $p\leq1$ are excluded by the wormhole defining equation (\ref{22}).

\subsubsection{\bf Case  $p<0$:}

It is obvious that this case is not supported by Eq.(\ref{22}), but if we
take an open universe $k=-1$ then this equation casts in the following form
\footnote{The case $p=0$ is not physically viable, so we have not considered this case throughout the paper.} 
\begin{equation}\label{22'}
\frac{\dot{a}^2}{a^2}=\frac{-1}{a^2}-\frac{3A^2}{2pa^{2p}}.
\end{equation}
Hence, we can allow $p$ to take the negative values and Eq.(\ref{22'}) is
written as
\begin{equation}\label{22''}
\frac{\dot{a}^2}{a^2}=\frac{3A^2}{2|p|}a^{2|p|}-\frac{1}{a^2}.
\end{equation}
This equation is describing wormhole solutions because it gives a throat
at $\dot{a}=0$ and $\dot{a}^2$ remains positive at large $a$. Therefore,
the wormhole solutions are obtained as
\begin{equation}
\tau-\tau_0=\int\frac{da}{\sqrt{\frac{3A^2}{2|p|}a^{2|p|+2}-1}},
\end{equation}
\begin{equation}
\phi(\tau)=\int Aa(\tau)^{|p|}d\tau,
\end{equation}
with a throat size at $\dot{a}=0$
\begin{equation}
a_0=\left({\frac{2|p|}{3A^2}}\right)^{\frac{1}{2|p|+2}}.
\end{equation}

\subsection{Case: $V(\phi)<0$ }

In this case, using the ansatz (\ref{8}) in  Eq.(\ref{21}) we find $p>3$
and the system of equations (\ref{19}), (\ref{20}) lead to the wormhole solutions
\begin{equation}\label{41}
\tau-\tau_0=\int\frac{da}{\sqrt{1-\frac{3A^2}{2pa^{2p-2}}}},
\end{equation}
\begin{equation}\label{42}
\phi(\tau)=\int\frac{A}{[a(\tau)]^p}d\tau,
\end{equation}
with a throat size at $\dot{a}=0$
\begin{equation}\label{43}
a_0=\left({\frac{3A^2}{2p}}\right)^{\frac{1}{2p-2}}.
\end{equation}

\subsection{General Case: $\dot{\phi}=g(a)$, $k=\pm1$}

We now consider the general case by assuming the ansatz \cite{Cost} and $k=\pm1$
\begin{equation}\label{31}
\dot{\phi}=g(a).
\end{equation}
Equation (\ref{20}) becomes
\begin{equation}\label{32}
g'\dot{a}+3\frac{\dot{a}}{a}g=-\frac{dV}{d\phi},
\end{equation}
where a prime denotes $d/da$. Using (\ref{31}) in the form $\dot{a}=g\frac{da}{d\phi}$, we obtain
\begin{eqnarray}\label{33}
V(a)&=&-\int\left[gg'+\frac{3g^2}{a}\right]da \\ \nonumber
&=&c_0-\frac{1}{2}g^2-3g^2\ln a+6\int g g' \ln a da,
\end{eqnarray}
where an integration by part has been used and $c_0$ is the constant of integration. Putting this form of potential into equation (\ref{19}) gives
\begin{equation}\label{34}
\frac{\dot{a}^2}{a^2}=\frac{k}{a^2}+3\int\frac{g^2}{a}da,
\end{equation}
or
\begin{equation}\label{35}
\tau-\tau_0=\int\frac{d{{a}}}{\left[k+3{{{a}}}^2\left[\int\frac{g^2}{{{a}}}d{{a}}\right]\right]^{1/2}},
\end{equation}
where $\tau_0$ as the integration constant denotes the time at which $a$
becomes $a_0\equiv a_{min}$. Notice that equations (\ref{34}), (\ref{35}) are different from the corresponding ones obtained in \cite{Cost} due to
the absence of a $g^2$ term (because of the previously mentioned
missing minus sign). If we demand the wormhole solutions of the form
\begin{equation}\label{36}
a=a_0+\sinh\frac{\tau-\tau_0}{m},
\end{equation}
where $m$ is a constant, then comparing (\ref{36}) with (\ref{35}) results
in
\begin{equation}\label{37}
g(a)=\left[\frac{2}{3}\left(\frac{aa_0-a_0^2+m^2-k}{m^2a^2}\right)\right]^{\frac{1}{2}}.
\end{equation}
Substituting (\ref{37}) into (\ref{33}) gives the potential in terms of the radius $a$ 
\begin{equation}\label{38}
V(a)=c_0+\frac{c_1}{a}+\frac{c_2}{a^2}.
\end{equation}
The scalar field as a function of $a$ can also be obtained by using (\ref{31}) and (\ref{35}) 
\begin{eqnarray}\label{39}
\phi(a)=\int \frac{g(a)da}{\left[k+3{{{a}}}^2\left[\int\frac{g^2}{{{a}}}d{{a}}\right]\right]^{1/2}},
\end{eqnarray}
where $g(a)$ is given by (\ref{37}). 

\section{Euclidean wormholes with minimally coupled Phantom field and perfect
fluid}

The action for the Phantom field minimally coupled to gravity and a perfect
fluid source is given by \cite{Sami}
\begin{equation}\label{44}
{\cal S}=\int d^4x \sqrt{-g}\left[{\cal R}+\frac{1}{2}\nabla_{\mu}\phi \nabla^{\mu}\phi-V(\phi)+{\cal
L}_{matter}\right].
\end{equation}
The energy momentum tensor and the corresponding Einstein equations are given by
\begin{eqnarray}\label{45}
T_{\mu \nu}=(\rho_b-p_b)u_{\mu}u_{\nu}+p_b g_{\mu \nu}-\nabla_{\mu}\phi \nabla_{\nu}\phi+g_{\mu
\nu}\left[\frac{1}{2}\nabla_{\lambda}\phi\nabla^{\lambda}\phi-V(\phi)\right],
\end{eqnarray}
\begin{equation}\label{46}
R_{\mu \nu}=(\rho_b-p_b)u_{\mu}u_{\nu}-\nabla_{\mu}\phi \nabla_{\nu}\phi-\frac{1}{2} g_{\mu \nu}(\rho_b+p_b-2V(\phi)).
\end{equation}
The Ricci tensor may be negative in Euclidean space if and only if the following condition is satisfied
\begin{equation}\label{47}
(\rho_b-p_b)u_{\mu}u_{\nu}-\nabla_{\mu}\phi \nabla_{\nu}\phi-\frac{1}{2} g_{\mu \nu}(\rho_b+p_b-2V(\phi))<0,
\end{equation}
where the potential can take the following cases as before
\begin{equation}\label{48}
 \left\{ \begin{array}{ll} V(\phi)\geq 0, 
 \\V(\phi)< 0.
 \\
\end{array}
\right.
\end{equation}
One may define the pressure and energy density of the Phantom field $\phi$ in Euclidean
signature 
\begin{equation}\label{49}
\rho_{\phi}=\frac{\dot{\phi}^2}{2}+V(\phi), \,\,\,\,p_{\phi}=\frac{\dot{\phi}^2}{2}-V(\phi). \end{equation}
Therefore, the conditions (\ref{48}) are rewritten as
\begin{equation}\label{50}
 \left\{ \begin{array}{ll} \rho_{\phi}\geq p_{\phi}, 
 \\\rho_{\phi}< p_{\phi}.
 \\
\end{array}
\right.
\end{equation}
To justify that the condition (\ref{47}) may be generally satisfied, one may for instance take a monotonic function $\phi(\tau)$ for which we obtain
\begin{equation}\label{51}
R_{00}=\frac{1}{2}(\rho_b+\rho_{\phi}-3p_b-p_{\phi})-\dot{\phi}^2.
\end{equation}
It is easily seen that $R_{00}$ may become negative provided that some suitable equations of state for matter and Phantom field are taken so that either
the terms including the energy densities and pressures become negative, or otherwise, the (Euclidean) kinetic term precedes these terms.
Unlike the pure Phantom field case where the wormholes are possible just for the closed and open universe, in the present model which includes the Phantom field accompanied by matter source we shall examine the existence of wormholes for the three cases of closed, flat and open universes.

\subsection{\bf Case $k=1$:}

The field equations in this case are obtained 
\begin{equation}\label{52}
\frac{\dot{a}^2}{a^2}=\frac{1}{a^2}-\left[\rho_b+\rho_{\phi}\right],
\end{equation}
\begin{equation}\label{53}
\ddot{\phi}+3\frac{\dot{a}}{a}\dot{\phi}=-\frac{dV(\phi)}{d\phi},
\end{equation}
where the background energy density due to the matter and radiation is given
by
\begin{equation}\label{54}
\rho_b=\frac{\rho_m}{a^3}+\frac{\rho_r}{a^4}.
\end{equation}
\\
If we limit ourselves to the ansatz (\ref{8}), and use (\ref{21}) and (\ref{49}) in (\ref{52}) we obtain
\begin{equation}\label{55}
\frac{\dot{a}^2}{a^2}=\frac{1}{a^2}-\left[\frac{\rho_m}{a^3}+\frac{\rho_r}{a^4}
+\frac{3A^2}{2pa^{2p}}\right],
\end{equation}
It is seen that $\dot{a}^2$ remains positive at large $a$ provided that $p>0$
which includes both conditions in (\ref{48}) or (\ref{50}). Comparing (\ref{22}) with (\ref{55}) indicates that the introduction of matter field, as a perfect
fluid with positive energy densities, changes the necessary condition for the occurrence of wormholes from $p>1$ into $p>0$, respectively.
Therefore, the wormhole solutions are obtained
\begin{equation}
\tau-\tau_0=\int\frac{da}{\sqrt{1-\left(\frac{\rho_m}{a}+\frac{\rho_r}{a^2}
+\frac{3A^2}{2pa^{2p-2}} \right)}},
\end{equation}
\begin{equation}
\phi(\tau)=\int\frac{A}{[a(\tau)]^p}d\tau,
\end{equation}
where the throat size at $\dot{a}=0$ is given by the solution of the following equation
\begin{equation}\label{56}
2p(a^{2p-2}-\rho_m a^{2p-3}-\rho_r a^{2p-4})-3A^2=0,
\end{equation}
which depends on the properties of the matter sources, namely $\rho_m$ and
$\rho_r$.

\subsection{\bf Case $k=0$:}

The field equations are obtained 
\begin{equation}\label{57}
\frac{\dot{a}^2}{a^2}=-\left[\frac{\rho_m}{a^3}+\frac{\rho_r}{a^4}
+\frac{3A^2}{2pa^{2p}}\right],
\end{equation}
\begin{equation}\label{58}
\ddot{\phi}+3\frac{\dot{a}}{a}\dot{\phi}=-\frac{dV(\phi)}{d\phi},
\end{equation}
It is easily seen that for positive values of $p$ we can
not cast the equation (\ref{57}) in the form of wormhole equation (\ref{11})
. In other words, $\dot{a}^2$ can not be positive for any positive value of $p$. Therefore, positive values of $p$ are excluded to be candidates for wormhole solutions. However, for the negative values of $p$ one can rewrite
(\ref{57}) as
\begin{equation}\label{59}
\frac{\dot{a}^2}{a^2}=\frac{3A^2a^{2|p|}}{2|p|}-\left[\frac{\rho_m}{a^3}+\frac{\rho_r}{a^4}
\right].
\end{equation}
This equation is describing wormhole solutions since $\dot{a}^2$ remains positive at large $a$ and results in asymptotically Euclidean wormholes 
\begin{equation}
\tau-\tau_0=\int\frac{da}{\sqrt{\frac{3A^2}{2|p|}a^{2|p|+2}-\left[\frac{\rho_m}{a}+\frac{\rho_r}{a^2}
\right]}},
\end{equation}
\begin{equation}
\phi(\tau)=\int Aa(\tau)^{|p|}d\tau,
\end{equation}
whose typical throat size at $\dot{a}=0$ is given by 
\begin{equation}\label{60}
\frac{3A^2}{2|p|}a^{2|p|+4}-\rho_m a-\rho_r=0.
\end{equation}

\subsection{\bf Case $k=-1$:}

The field equations are given by 
\begin{equation}\label{61}
\frac{\dot{a}^2}{a^2}=-\frac{1}{a^2}-\left[\frac{\rho_m}{a^3}+\frac{\rho_r}{a^4}
+\frac{3A^2}{2pa^{2p}}\right],
\end{equation}
\begin{equation}\label{62}
\ddot{\phi}+3\frac{\dot{a}}{a}\dot{\phi}=-\frac{dV(\phi)}{d\phi}.
\end{equation}
The equation (\ref{61}), for positive values of $p$, does not describe wormhole solutions
since $\dot{a}^2$ can never be positive.
Therefore, the positive values of $p$ are again excluded from consideration. However, for negative values of $p$ one can rewrite (\ref{61}) as
\begin{equation}\label{63}
\frac{\dot{a}^2}{a^2}=\frac{3A^2a^{2|p|}}{2|p|}-\left[\frac{1}{a^2}+\frac{\rho_m}{a^3}+\frac{\rho_r}{a^4}
\right].
\end{equation}
Similar to (\ref{59}), the equation (\ref{63}) may describe wormhole solutions
\begin{equation}
\tau-\tau_0=\int\frac{da}{\sqrt{\frac{3A^2}{2|p|}a^{2|p|+2}-\left[1+\frac{\rho_m}{a}+\frac{\rho_r}{a^2}
\right]}},
\end{equation}
\begin{equation}
\phi(\tau)=\int Aa(\tau)^{|p|}d\tau,
\end{equation}
whose throat size is given by the solution of the following equation
\begin{equation}\label{64}
\frac{3A^2}{2|p|}a^{2|p|+4}-a^2-\rho_m a-\rho_r=0.
\end{equation}

\subsection{\bf General Case: $\dot{\phi}=g(a)$, $k=\pm1, 0$}

We again consider the general case by assuming the ansatz (\ref{31}). Moreover,
we shall consider the three cases of closed, flat and open universes indicated
by $k=\pm1, 0$. In the same way as discussed in the subsection 3.3, we obtain the general field equation
\begin{equation}\label{65}
\frac{\dot{a}^2}{a^2}=\frac{k}{a^2}-\frac{\rho_m}{a^3}-\frac{\rho_r}{a^4}+3\int\frac{g^2}{a}da,
\end{equation}
or
\begin{equation}\label{66}
\tau-\tau_0=\int\frac{d{{a}}}{\left[k-\frac{\rho_m}{a}-\frac{\rho_r}{a^2}+3{{{a}}}^2\left[\int\frac{g^2}{{{a}}}d{{a}}\right]\right]^{1/2}}.
\end{equation}
Demanding the wormhole solutions of the form (\ref{36}) leads to
\begin{equation}\label{67}
g(a)=\left[\frac{2}{3}\left(\frac{aa_0-a_0^2+m^2-k}{m^2a^2}\right)-\frac{1}{3}\left(
\frac{3\rho_m}{a^3}+\frac{4\rho_r}{a^4}\right)\right]^{\frac{1}{2}},
\end{equation}
\begin{equation}
V(a)=c_0+\frac{c_1}{a}+\frac{c_2}{a^2}+\frac{c_3}{a^3}+\frac{c_4}{a^4}.
\end{equation}
and
\begin{eqnarray}
\phi(a)=\int \frac{g(a)da}{\left[k-\frac{\rho_m}{a}-\frac{\rho_r}{a^2}+3{{{a}}}^2\left[\int\frac{g^2}{{{a}}}d{{a}}\right]\right]^{1/2}}.
\end{eqnarray}

\section{Conclusion}

The possible forms of matter sources which may result in classical Euclidean wormholes are very limited. The axion field, conformal scalar field, (pure imaginary) minimally coupled scalar field, conformally coupled scalar field, a third rank antisymmetric tensor field and some special kinds of matter sources result in such wormhole solutions. However, pure gravity or a (real) minimally coupled scalar field do not represent Euclidean wormholes. Any effort to obtain new forms of matter representing Euclidean wormhole solutions are of particular importance in quantum gravity. In this paper, we have considered the Phantom field minimally coupled to gravity and studied the possibility of Euclidean wormholes to occur in the spatially closed and open Friedmann-Robertson-Walker
universes. It is shown that these solutions may appear due to the negative kinetic energy of the Phantom field. Then, we have obtained some wormhole solutions in this model and found the general
form of the corresponding Phantom field potential. Then, we have studied
the possibility of Euclidean wormholes to occur in the system of a Phantom field accompanied by perfect fluid, minimally coupled to gravity. The existence
of these solutions is explicitly shown and wormhole solutions together with the corresponding Phantom field potentials are obtained for spatially closed, flat and open Friedmann-Robertson-Walker universes. 

It is appealing to study other models in which the kinetic term may effectively
get wrong sign similar to the Phantom field. Kinetically driven inflation
or {\it k-Inflation} models are examples of this kind where a large class of higher-order (i.e. non-quadratic) scalar kinetic terms can, without the help of potential terms, drive an inflationary evolution starting from rather generic initial conditions \cite{k-inf}. We aim to study the possibility of obtaining
Euclidean wormholes in such models as well.

\section*{Acknowledgment}
This work  has been supported by the Research office of Azarbaijan
University of Tarbiat Moallem, Tabriz, Iran.


\newpage

\begin{thebibliography}{99}
\bibitem{Haw}S. W. Hawking, Phys. Rev. D.{\bf 37} (1988), 904.
\bibitem{Col1}S. Coleman, Nucl. Phys. B{\bf 307} (1988), 864.
\bibitem{Col2}S. Coleman, Nucl. Phys. B{\bf 310} (1988), 643.
\bibitem{Col3}I. Klebanov, L. Susskind, and T. Banks, Nucl. Phys. B{\bf 317} (1989), 665.
\bibitem{Mig}S. Mignemi and I. Moss, Phys. Rev. D.{\bf 48} (1993), 3725.
\bibitem{Chee}J. Cheeger and D. Grommol, Ann. Math. {\bf 96} (1972), 413.
\bibitem{Gid1}S. Giddings, A. Strominger, Nucl. Phys. B{\bf 306}
(1988), 890.
\bibitem{Gid2}J. J. Halliwell and R. Laflamme, Class. Quantum. Grav. {\bf
6} (1989), 1839.
\bibitem{Sahni1}E. J. Copeland, M. Sami, and S. Tsujikawa, Int. J. Mod. Phys.D{\bf
15}, (2006), 1753.
\bibitem{Gid3}K. Lee, Phys. Rev. Lett.{\bf 61} (1988), 263.
\bibitem{Gid4}A. Hosoya, W. Ogura, Phys. Lett. B{\bf 225} (1989), 117.
\bibitem{Gid5}A. K. Gupta, J. Hughes, J. Preskill, and M. B. Wise, Nucl. Phys. B{\bf 333} (1990), 195.
\bibitem{Hoyl}F. Hoyle, Mon. Not. R. Astr. Soc. {\bf 108}, (1948), 372; Mon. Not. R. Astr. Soc. {\bf 109}, (1949), 365.
\bibitem{HN}F. Hoyle and J. V. Narlikar, Proc. Roy. Soc. A{\bf 282}, 
(1964), 191; Mon. Not. R. Astr. Soc. {\bf 155}, (1972), 305.
\bibitem{Sahni}V. Sahni and Y. Shtanov, JCAP {\bf 0311},(2003), 014;
V. Sahni and Y. Shtanov, Int. J. Mod. Phys. D.{\bf 67}, (2000), 1515;
E. Elizalde, S. Nojiri and S. D. Odintsov, Phys. Rev. D.{\bf 70}, (2004),
043539.
\bibitem{Cald}R. R. Caldwell, Phys. Lett. B.{\bf 545}, (2002), 23; R. R. Caldwell {\it et al}, Phys. Rev. Lett. 91, (2003), 071301.
\bibitem{Sbrane}C. M. Chen, D. V. Gal'tsov and M. Gutperle, Phys.
Rev. D.{\bf 66}, (2002),  024043 ; P. K. Townsend and
M. N. R. Wohlfarth, Phys. Rev. Lett.{\bf 91}, (2003), 061302;
N. Ohta, Phys. Lett. B.{\bf 558}, (2003), 213; Phys. Rev.
Lett.{\bf 91}, (2003), 061303; Prog. Theor. Phys.{\bf 110},
(2003), 269; Int. J. Mod. Phys. A.{\bf 20}, (2005), 1; S. Roy, Phys.
Lett. B.{\bf 567}, (2003), 322.
\bibitem{Cost}S. Costakis, P. Leach, and G. Flessa, Phys. Rev. D.{\bf 49} (1994), 6489.
\bibitem{HP}S. W. Hawking, and D. Page, Phys. Rev. D.{\bf 42} (1990), 2655.
\bibitem{Coule}D. H. Coule, Phys. Rev. D.{\bf 55} (1997), 6606.
\bibitem{Wald}R. Wald, {\it General Relativity} (University of Chicago Press,
Chicago, 1984).
\bibitem{Sami}M. Sami, A. Toporensky, Mod. Phys. Lett. A.{\bf 19}, (2004),
1509.
\bibitem{k-inf}C. Armend\`ariz-Pic\`ona, T. Damour, and V. Mukhanov, Phys.
Lett. B.{\bf 458}, (1999), 209.
\end{thebibliography}
\end{document}